\pdfoutput=1
\documentclass[11pt]{article}
\usepackage[pdftex]{graphicx,color} 
\usepackage{jheppub}
\usepackage{amsmath}
\usepackage{amssymb}
\usepackage{multirow}
\usepackage{mathtools}
\usepackage{dsdshorthand}
\usepackage[utf8]{inputenc}
\newcommand{\bea}{\begin{equation}\begin{aligned}}
\newcommand{\eea}[1]{\label{#1}\end{aligned}\end{equation}}
\newcommand{\beq}{\begin{equation}}
\newcommand{\eeq}{\end{equation}}

\usepackage{tikz}
\usetikzlibrary{arrows,calc,shapes,decorations.pathmorphing,positioning}
\tikzset{
>=stealth',
help lines/.style={dashed, thick},
axis/.style={<->},
important line/.style={thick},
connection/.style={thick, dotted},
  cross/.style={
    cross out,
    draw=black, 
    minimum size=7pt, 
    inner sep=0pt,
    outer sep=0pt
  },
  branchcut/.style={
    decoration={
      snake,
      amplitude=1pt,
      segment length=6pt,
    },
    decorate,
    thick
  },
}

\title{Analytic Bootstrap for Boundary CFT}
\author{Agnese Bissi,}
\author{Tobias Hansen,}
\author{Alexander Söderberg}
\affiliation{Department of Physics and Astronomy,
	Uppsala University,\\
	Box 516,
	SE-751 20 Uppsala,
	Sweden}
\emailAdd{agnese.bissi@physics.uu.se, tobias.hansen@physics.uu.se, alexander.soderberg@physics.uu.se}

\abstract{We propose a method to analytically solve the bootstrap equation for two point functions in boundary CFT. We consider the analytic structure of the correlator in Lorentzian signature and in particular the discontinuity of bulk and boundary conformal blocks to extract CFT data.
As an application, the correlator $\< \f \f \>$ in $\f^4$ theory at the Wilson-Fisher fixed point is computed to order $\e^2$ in the $\e$ expansion.}

\begin{document}
\maketitle

\section{Introduction}

Over the last ten years much progress has been made in understanding the dynamics of Conformal Field Theories (CFT) in dimensions greater than two, using both analytical and numerical conformal bootstrap techniques. They rely on the formulation of consistency conditions on conformal dimensions and three point function coefficients (CFT data) of local primary operators, arising from the associativity of the Operator Product Expansion (OPE) and symmetries of the theory. Analytic solutions of crossing equations are not easy to obtain. Recently, two equivalent approaches have been proposed to analytically extract CFT data. In one approach, the main observation has been that specific singularities of the four point correlator completely fix the large spin expansion of the CFT data, making it possible to reconstruct the CFT data even for finite spin \cite{Alday:2016njk}. In the other approach, CFT data can be derived as an integral of the double discontinuity of the four point correlator over Minkowski regions \cite{Caron-Huot:2017vep}. In the latter, the structure of the singularities of the correlator in Lorentzian signature plays an important role.

Local operators are not enough to completely cover the set of observables in a generic CFT. The study of extended objects, such as conformal defects or boundaries, complements the information which can be extracted from bulk correlation functions, in addition to naturally arising in experimental setups. In particular the rich interplay between the dynamics of fields living in the bulk and on the defect is completely inaccessible from the analysis of bulk field correlation functions only. Thus the CFT data is enlarged to accommodate the conformal dimensions of defect operators and the
Boundary Operator Expansion (BOE) coefficients governing the expansion of bulk operators in
terms of boundary operators.
Lately, defects in conformal field theories have received a lot of attention \cite{Billo:2016cpy, Gadde:2016fbj, Liendo:2016ymz, deLeeuw:2017dkd, Rastelli:2017ecj, Soderberg:2017oaa, Herzog:2017xha, Karch:2017wgy, Fukuda:2017cup, Sato:2017gla, Lauria:2017wav, Lemos:2017vnx, Goncalves:2018fwx, Prochazka:2018bpb, Bianchi:2018zpb, Kobayashi:2018okw, Karch:2018uft, Guha:2018snh, Isachenkov:2018pef, Liendo:2018ukf,Lauria:2018klo}. In this paper we will be interested in CFTs in the presence of boundaries (BCFT), which are conformal defects of codimension one. 

The bootstrap approach for such systems has been initiated in \cite{Liendo:2012hy},
using in parts the thorough treatment of BCFTs in \cite{McAvity:1995zd}. The main idea is to use either
the OPE between bulk operators or the BOE in the two point function of local scalar operators. Analogously to the case of four point functions, the compatibility of these two expansions results in the bootstrap equation which constrains also the boundary CFT data. In \cite{Liendo:2012hy}, an analytic solution to this equation for the correlator $\< \f \f \>$ in the Wilson-Fisher model has been found to order $\epsilon$. In this paper we are going to extend this result and provide analytical CFT data to order  $\epsilon^2$. The obstacle in extracting this CFT data is the fact that at this order in $\epsilon$, there are infinitely many operators appearing both in the bulk and in the boundary channel expansions. 

The method that we are using relies on the analytic structure of both bulk and boundary blocks. The crucial observation is that for specific values of the dimensions of intermediate operators, the branch cut structure of the blocks dramatically simplifies, allowing to reduce the problem
from two to one infinite sum of blocks.
This enables us to find consistency relations for the OPE coefficients and the anomalous dimensions in both channels up to order $\epsilon^2$. As a check of our results, we verified that the anomalous dimensions, which are already known in the literature, satisfy the relations. Giving the anomalous dimensions and the structure of the OPE as an input, it is possible to compute to order $\epsilon^2$ the OPE coefficients and hence the full two point correlator, which is presented in Section \ref{sec:correlator}.

This approach is very similar in spirit to \cite{Caron-Huot:2017vep}, where the double discontinuity of the four point correlator in a homogeneous CFT is used to compute the OPE coefficients
and \cite{Lemos:2017vnx}, where the same idea was applied to defect CFTs (DCFT) with codimension greater than one.
However the case of BCFT is simpler: there is only one cross ratio, and the analytic structure of branch cuts is simpler than the CFT and DCFT cases, making it possible to invert the crossing equation and obtain CFT data. Another simplification is that our example involves only scalar operators. In this sense, the present paper provides a more accessible example for OPE inversion.

The organization of the paper is as follows. In Section \ref{sec:analytic_structure} we discuss the analytic structure of bulk and boundary blocks and review the bootstrap equation for the BCFT case. In Section \ref{sec:3} we review how to extract CFT data for the Wilson-Fisher BCFT to order $\epsilon$.
Section \ref{sec:4} contains the main results of this paper. We present how to compute CFT data to order $\epsilon^2$ using the analytic structure of the two point function and symmetries of the BCFT. We conclude with a discussion of other potential applications of the method we proposed and some future directions.


\section{Analytic structure of BCFT correlators}
\label{sec:analytic_structure}

\begin{figure}
\centering
  \begin{tikzpicture}[scale=1]
    \coordinate (n) at (0,3);
    \coordinate (e) at (3,0);
    \coordinate (w) at (-3,0);
    \coordinate (s) at (0,-3);
    \coordinate (bp1) at (1,0);
    \coordinate (bp2) at (-1,0);
    \draw[->] (w) --  (e) ;
    \draw[->] (s) --  (n) ;
    \draw [line width=1mm] (0,2.5) --  (0,-2.5) ;
    \filldraw [black] (2,1) circle (2pt) node[below right, black] {$x$};
    \filldraw [black] (1,-1) circle (2pt) node[below right, black] {$y$};
    \filldraw [black] (-2,1) circle (2pt) node[below left, black] {$\bar{x}$};
    \filldraw [black] (-1,-1) circle (2pt) node[below left, black] {$\bar{y}$};
    \node at (0,3) [left] {$\vec{x}$};
    \node at (3,0) [below] {$x_\perp$};
  \end{tikzpicture}
\caption{Coordinates $x$, $y$ and boundary at $x_\perp = 0$. Also pictured are the mirror images $\bar{x}$ and $\bar{y}$.}
\label{fig:coordinates}
\end{figure}
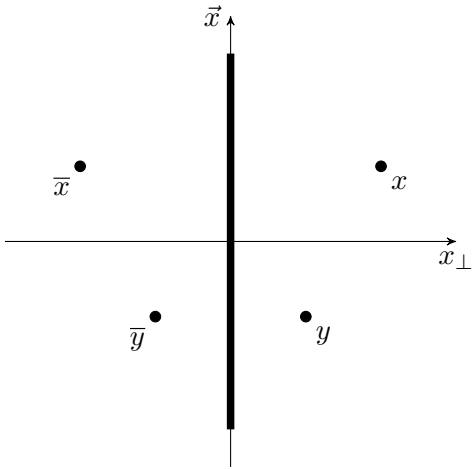

We study the two-point function of a scalar operator $\phi$
\beq
\< \phi (x) \phi(y) \> = \frac{F(z)}{(4 x_\perp y_\perp)^{\Delta_\phi}}\,,
\eeq
where the coordinates $x^\mu = (\vec{x},x_\perp)$ are split into
the $d-1$ coordinates tangential to the boundary $\vec{x}$ and
the distance from the boundary $x_\perp \geq 0$, as illustrated in Figure \ref{fig:coordinates}.
The correlator can be written in terms of a function of the single cross-ratio\footnote{We prefer this over the more conventional choice $\xi = z - \frac{1}{2}$ of \cite{McAvity:1995zd,Liendo:2012hy} because $z$ simply changes by a minus sign when replacing $x$ or $y$ by its mirror image on the other side of the boundary, e.g. $x_{\perp} \to -x_{\perp}$.}
\begin{equation}\label{defeta}
z= \frac{(\vec{x}-\vec{y})^2+x_{\perp}^2 +y_{\perp}^2}{4x_{\perp} y_{\perp}} \,.
\end{equation}
The function $F(z)$ can be expanded into conformal blocks in two different ways.\footnote{For a more detailed description of the expansions see \cite{Liendo:2012hy}.}
One can expand in boundary conformal blocks by expanding both operators
in terms of fields living on the boundary
\begin{align}
\phi(x) &= \sum_{\mathcal{O}}\frac{\mu_{\hat{\De}}}{x_\perp^{\Delta_{\phi} - \hat{\Delta}}}B_{\hat{\Delta}}(x_\perp^2, \vec{\partial}^2)\hat{\mathcal{O}}_{\hat{\Delta}}(\vec{x}) \, .
\end{align}
Here $\mu_{\hat{\De}}$ are the BOE coefficients, and $B_{\hat{\Delta}}(x_\perp^2, \vec{\partial}^2)$ are differential operators that generate descendants on the boundary \cite{McAvity:1995zd}.
Notice that scalar bulk operators are expanded into scalar operators on the boundary.
Then one uses that the two-point function of boundary operators is orthogonal to
obtain the expansion
\beq
F(z) = \sum_{\hat{\De}} \mu^2_{\hat{\De}} g_{i}(\hat{\Delta},z)\,,
\label{eq:boundary_expansion}
\eeq
where the boundary channel conformal block is given by\footnote{The subscript $i$ stands for interface.}
\bea
g_{i}(\hat{\Delta},z) &= (z-\tfrac{1}{2})^{- \hat{\Delta}} {}_2 F_1 \left(\hat{\Delta},\hat{\De} + 1 - \frac{d}{2}; 2 \hat{\De} + 2 - d; \frac{1}{\tfrac{1}{2} - z}\right)\, .
\eea{eq:gi}
Another possibility is to first take the usual OPE between the two operators in the bulk, yielding a sum over one-point functions.
In the presence of a boundary, not only the identity but also all other scalar fields (and only scalar fields) have a non-vanishing one-point function \cite{Billo:2016cpy}. In this way the correlator is expanded into bulk conformal blocks
\beq
F(z) = (z-\tfrac{1}{2})^{-\Delta_\phi} \sum_\De \lambda_\De a_\De g_{b}(\Delta,z)\,.
\label{eq:bulk_expansion}
\eeq
Here $\lambda_\De$ are the usual bulk OPE coefficients, $a_\De$ are the coefficients of bulk one-point functions and the bulk conformal blocks are given by
\bea
g_{b}(\Delta,z) &= (z-\tfrac{1}{2})^{\Delta / 2} {}_2 F_1\left(\frac{\De}{2},\frac{\De}{2};\De + 1 - \frac{d}{2};\tfrac{1}{2} - z\right)\, .
\eea{eq:gb}
In both cases the exchanged operators are scalars and labeled only by their conformal dimensions $\De$ or $\hat{\Delta}$.
The statement that both expansions \eqref{eq:boundary_expansion} and \eqref{eq:bulk_expansion} are equal is the bootstrap equation.
\begin{figure}
\centering
  \begin{tikzpicture}[scale=1]
    \coordinate (n) at (0,2.5);
    \coordinate (e) at (3,0);
    \coordinate (w) at (-3,0);
    \coordinate (s) at (0,-2.5);
    \coordinate (bp1) at (1,0);
    \coordinate (bp2) at (-1,0);
    \draw[->] (w) --  (e) ;
    \draw[->] (s) --  (n) ;
	\draw [branchcut] (bp2) -- (bp1);
    \node at (1,0) [cross] {};
    \node at (-1,0) [cross] {};
    \node at (1,-0.5) [] {$\frac{1}{2}$};
    \node at (-1,-0.5) [] {$-\frac{1}{2}$};
    \node at (2.7,2.2) [] {$z$};
    \draw[-] (2.5,2) --  (2.9,2) ;
    \draw[-] (2.5,2) --  (2.5,2.4) ;
  \end{tikzpicture}\qquad \qquad
  \begin{tikzpicture}[scale=1]
    \coordinate (n) at (0,2.5);
    \coordinate (e) at (3,0);
    \coordinate (w) at (-3,0);
    \coordinate (s) at (0,-2.5);
    \coordinate (bp1) at (1,0);
    \coordinate (bp2) at (-1,0);
    \draw[->] (w) --  (e) ;
    \draw[->] (s) --  (n) ;
	\draw [branchcut] (w) -- (bp2);
    \node at (1,0) [cross] {};
    \node at (-1,0) [cross] {};
    \node at (1,-0.5) [] {$\frac{1}{2}$};
    \node at (-1,-0.5) [] {$-\frac{1}{2}$};
    \node at (2.7,2.2) [] {$z$};
    \draw[-] (2.5,2) --  (2.9,2) ;
    \draw[-] (2.5,2) --  (2.5,2.4) ;
  \end{tikzpicture}

\caption{Analytic structure of $g_i(n,z)$ (left) and $g_b(2n,z)$ (right) for positive integer $n$.} \label{fig:analytic_structure_gi_gb}
\end{figure}
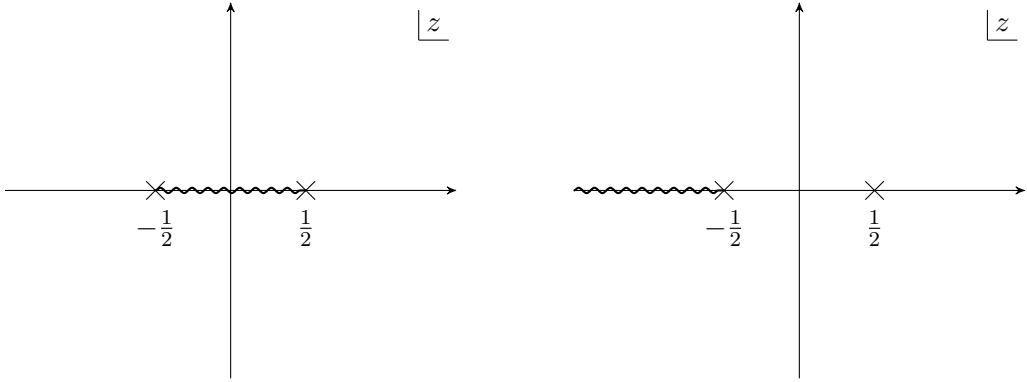

\begin{figure}
\centering
  \begin{tikzpicture}[scale=1]
    \coordinate (n) at (0,2.5);
    \coordinate (e) at (3,0);
    \coordinate (w) at (-3,0);
    \coordinate (s) at (0,-2.5);
    \coordinate (bp1) at (1,0);
    \coordinate (bp2) at (-1,0);
    \draw[->] (w) --  (e) ;
    \draw[->] (s) --  (n) ;
    \filldraw [black] 
     (2,0) circle (2pt) node[below right, black] {$z$};
	\draw [branchcut] (w) -- (bp1);
    \node at (1,0) [cross] {};
    \node at (-1,0) [cross] {};
    \node at (1,-0.5) [] {$\frac{1}{2}$};
    \node at (-1,-0.5) [] {$-\frac{1}{2}$};
    \draw[->] (2,0) arc (0:175:2);
    \draw[->] (2,0) arc (0:-175:2);
    \node at (-2.5,0.5) [] {$z e^{i \pi}$};
    \node at (-2.5,-0.5) [] {$z e^{-i \pi}$};
  \end{tikzpicture}

\caption{Analytic structure of $F(z)$ and paths of analytic continuation to negative $z$.} \label{fig:ac}
\end{figure}
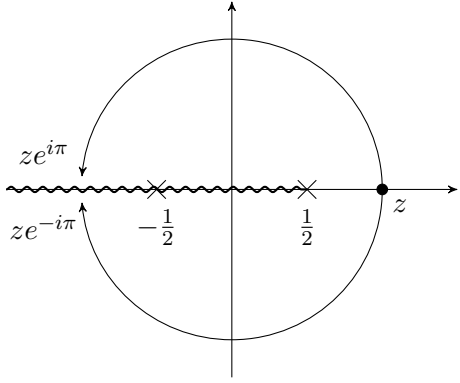

The analytic structure of the conformal blocks is simple. They both have singularities at $z = \pm \frac{1}{2}$ and $z = \infty$. There are branch cuts on the real axis for $z < \frac{1}{2}$ that arise from the different factors
\bea
&\text{function} &&\text{branch cut}\\
&(z-\tfrac{1}{2})^{a}\,, && z \in (-\infty, \tfrac{1}{2}) \text{ iff } a \notin \mathbb{Z}\,,\\
&{}_2 F_1 \left(a,b,c; (\tfrac{1}{2} - z)^{-1}\right)\,, \qquad && z \in (-\tfrac{1}{2}, \tfrac{1}{2})\,,\\
&{}_2 F_1 \left(a,b,c; \tfrac{1}{2} - z\right)\,, && z \in (-\infty, -\tfrac{1}{2})\,.\\
\eea{eq:branch_cuts}
An interesting observation is that whenever the exponents in the conformal blocks are integers, the blocks do not have a branch cut in the whole region $z < \frac{1}{2}$, but only the branch cut of the hypergeometric function (see Figure \ref{fig:analytic_structure_gi_gb}).
This can be turned into a powerful computational tool: By taking the discontinuity of the bootstrap equation at $z \in (-\infty, -\tfrac{1}{2})$, all boundary blocks for integer dimensions can be removed from the equation. Similarly, taking the discontinuity at $z \in (-\tfrac{1}{2}, \tfrac{1}{2})$ removes all bulk blocks for even integer dimensions.
Later on this is what will allow us to solve the bootstrap at order $\epsilon^2$. The bootstrap equation will have infinite sums on both sides. One of them can be removed by taking the discontinuity.
In practice we will only use the discontinuity that removes boundary blocks for integer dimensions
\beq
\text{Disc } F(z) \equiv F(z e^{i \pi}) - F(z e^{-i \pi}) \,, \qquad z \in (\tfrac{1}{2},+\infty)\,.
\eeq
The two paths of analytic continuation are illustrated in Figure \ref{fig:ac}.
One might wonder whether the BOE \eqref{eq:boundary_expansion} and the OPE \eqref{eq:bulk_expansion} are still convergent when doing these analytic continuations.
This question can be answered by considering radial coordinates, which were introduced for DCFT in \cite{Lauria:2017wav}.\footnote{We thank Marco Meineri for suggesting this.} For the special case of boundary CFTs they are given by
\beq
\hat{r} (z) = 2 \left(z - \sqrt{(z+\tfrac{1}{2})(z-\tfrac{1}{2})} \right)\,, \qquad
r (z) = \frac{z+\frac{3}{2}-2 \sqrt{z + \frac{1}{2}}}{z-\frac{1}{2}}\,.
\eeq
These coordinates encode the region of convergence for the two expansions we are using
\bea
|\hat{r}(z)| &< 1 \,, &\qquad& \text{region of BOE convergence}\,,\\
|r(z)| &< 1 \,, &\qquad& \text{region of OPE convergence}\,.
\eea{eq:convergence_regions}
After analytic continuation these coordinates become
\bea
\hat{r} (z e^{\pm i \pi}) &= -\hat{r} (z) \,,\\
r (z e^{\pm i \pi}) &= \frac{z-\frac{3}{2}+2 \sqrt{z - \frac{1}{2}} e^{\pm \frac{i \pi}{2}}}{z+\frac{1}{2}}
\quad \Rightarrow \quad |r (z e^{\pm i \pi})| = 1\,, \quad \forall z \in (\tfrac{1}{2},+\infty)\,.
\eea{eq:radial_coords_ac}
We conclude that $\hat{r}$ just changes sign and the convergence of the BOE after analytic continuation is ensured. The coordinate $r$ however approaches the boundary of its region of convergence at the end of the analytic continuation.
This means that we have to be cautious when commuting analytic continuation and the OPE sum. In our example below we will explicitly check that they commute.

\begin{figure}
\centering
  \begin{tikzpicture}[scale=.75]
    \coordinate (n) at (0,3);
    \coordinate (e) at (3,0);
    \coordinate (w) at (-3,0);
    \coordinate (s) at (0,-3);
    \coordinate (bp1) at (1,0);
    \coordinate (bp2) at (-1,0);
    \draw[->] (w) --  (e) ;
    \draw[->] (s) --  (n) ;
    \node at (0,3) [left] {$t$};
    \node at (3,0) [below] {$x_\perp$};
    \node at (-3.2,-3) [above right] {$z>\tfrac{1}{2}$};
    \node at (-3.2,-3.7) [above right] {$(x-y)^2 > 0$};
    \node at (-3.2,-4.4) [above right] {$(\bar{x}-y)^2 > 0$};
    \draw [line width=1mm] (0,2.5) --  (0,-2.5) ;
    \filldraw [black] (2,.3) circle (2pt) node[right, black] {$x$};
    \node (x) at (2,.3) [] {};
    \filldraw [black] (.8,-.3) circle (2pt) node[right, black] {$y$};
    \node (y) at (.8,-.3) [] {};
    \filldraw [black] (-2,.3) circle (2pt) node[left, black] {$\bar{x}$};
    \node (xb) at (-2,.3) [] {};
    \filldraw [black] (-.8,-.3) circle (2pt) node[left, black] {$\bar{y}$};
    \node (yb) at (-.8,-.3) [] {};
    \begin{scope}[node distance=1]
    \node[above right = of x] (xne) {};
    \node[below right = of x] (xse) {};
    \node[below left = of x] (xsw) {};
    \node[above left = of x] (xnw) {};
    \draw[densely dotted] (xne) --  (xsw);
    \draw[densely dotted] (xnw) --  (xse);
    \node[above right = of y] (yne) {};
    \node[below right = of y] (yse) {};
    \node[below left = of y] (ysw) {};
    \node[above left = of y] (ynw) {};
    \draw[densely dotted] (yne) --  (ysw);
    \draw[densely dotted] (ynw) --  (yse);
    \node[above right = of xb] (xbne) {};
    \node[below right = of xb] (xbse) {};
    \node[below left = of xb] (xbsw) {};
    \node[above left = of xb] (xbnw) {};
    \draw[densely dotted] (xbne) --  (xbsw);
    \draw[densely dotted] (xbnw) --  (xbse);
    \node[above right = of yb] (ybne) {};
    \node[below right = of yb] (ybse) {};
    \node[below left = of yb] (ybsw) {};
    \node[above left = of yb] (ybnw) {};
    \draw[densely dotted] (ybne) --  (ybsw);
    \draw[densely dotted] (ybnw) --  (ybse);
    \end{scope}
  \end{tikzpicture}
  \begin{tikzpicture}[scale=0.75]
    \coordinate (n) at (0,3);
    \coordinate (e) at (3,0);
    \coordinate (w) at (-3,0);
    \coordinate (s) at (0,-3);
    \coordinate (bp1) at (1,0);
    \coordinate (bp2) at (-1,0);
    \draw[->] (w) --  (e) ;
    \draw[->] (s) --  (n) ;
    \node at (0,3) [left] {$t$};
    \node at (3,0) [below] {$x_\perp$};
    \node at (-3.2,-3) [above right] {$-\tfrac{1}{2}<z<\tfrac{1}{2}$};
    \node at (-3.2,-3.7) [above right] {$(x-y)^2 < 0$};
    \node at (-3.2,-4.4) [above right] {$(\bar{x}-y)^2 > 0$};
    \draw [line width=1mm] (0,2.5) --  (0,-2.5) ;
    \filldraw [black] (1.6,.5) circle (2pt) node[right, black] {$x$};
    \node (x) at (1.6,.5) [] {};
    \filldraw [black] (1.3,-.5) circle (2pt) node[right, black] {$y$};
    \node (y) at (1.3,-.5) [] {};
    \filldraw [black] (-1.6,.5) circle (2pt) node[left, black] {$\bar{x}$};
    \node (xb) at (-1.6,.5) [] {};
    \filldraw [black] (-1.3,-.5) circle (2pt) node[left, black] {$\bar{y}$};
    \node (yb) at (-1.3,-.5) [] {};
    \begin{scope}[node distance=1.2]
    \node[above right = of x] (xne) {};
    \node[below right = of x] (xse) {};
    \node[below left = of x] (xsw) {};
    \node[above left = of x] (xnw) {};
    \draw[densely dotted] (xne) --  (xsw);
    \draw[densely dotted] (xnw) --  (xse);
    \node[above right = of y] (yne) {};
    \node[below right = of y] (yse) {};
    \node[below left = of y] (ysw) {};
    \node[above left = of y] (ynw) {};
    \draw[densely dotted] (yne) --  (ysw);
    \draw[densely dotted] (ynw) --  (yse);
    \node[above right = of xb] (xbne) {};
    \node[below right = of xb] (xbse) {};
    \node[below left = of xb] (xbsw) {};
    \node[above left = of xb] (xbnw) {};
    \draw[densely dotted] (xbne) --  (xbsw);
    \draw[densely dotted] (xbnw) --  (xbse);
    \node[above right = of yb] (ybne) {};
    \node[below right = of yb] (ybse) {};
    \node[below left = of yb] (ybsw) {};
    \node[above left = of yb] (ybnw) {};
    \draw[densely dotted] (ybne) --  (ybsw);
    \draw[densely dotted] (ybnw) --  (ybse);
    \end{scope}
  \end{tikzpicture}
  \begin{tikzpicture}[scale=0.75]
    \coordinate (n) at (0,3);
    \coordinate (e) at (3,0);
    \coordinate (w) at (-3,0);
    \coordinate (s) at (0,-3);
    \coordinate (bp1) at (1,0);
    \coordinate (bp2) at (-1,0);
    \draw[->] (w) --  (e) ;
    \draw[->] (s) --  (n) ;
    \node at (0,3) [left] {$t$};
    \node at (3,0) [below] {$x_\perp$};
    \node at (-3.2,-3) [above right] {$z<-\tfrac{1}{2}$};
    \node at (-3.2,-3.7) [above right] {$(x-y)^2 < 0$};
    \node at (-3.2,-4.4) [above right] {$(\bar{x}-y)^2 < 0$};
    \draw [line width=1mm] (0,2.5) --  (0,-2.5) ;
    \filldraw [black] (.6,1) circle (2pt) node[right, black] {$x$};
    \node (x) at (.6,1) [] {};
    \filldraw [black] (.3,-.8) circle (2pt) node[right, black] {$y$};
    \node (y) at (.3,-.8) [] {};
    \filldraw [black] (-.6,1) circle (2pt) node[left, black] {$\bar{x}$};
    \node (xb) at (-.6,1) [] {};
    \filldraw [black] (-.3,-.8) circle (2pt) node[left, black] {$\bar{y}$};
    \node (yb) at (-.3,-.8) [] {};
    \begin{scope}[node distance=1.2]
    \node[above right = of x] (xne) {};
    \node[below right = of x] (xse) {};
    \node[below left = of x] (xsw) {};
    \node[above left = of x] (xnw) {};
    \draw[densely dotted] (xne) --  (xsw);
    \draw[densely dotted] (xnw) --  (xse);
    \node[above right = of y] (yne) {};
    \node[below right = of y] (yse) {};
    \node[below left = of y] (ysw) {};
    \node[above left = of y] (ynw) {};
    \draw[densely dotted] (yne) --  (ysw);
    \draw[densely dotted] (ynw) --  (yse);
    \node[above right = of xb] (xbne) {};
    \node[below right = of xb] (xbse) {};
    \node[below left = of xb] (xbsw) {};
    \node[above left = of xb] (xbnw) {};
    \draw[densely dotted] (xbne) --  (xbsw);
    \draw[densely dotted] (xbnw) --  (xbse);
    \node[above right = of yb] (ybne) {};
    \node[below right = of yb] (ybse) {};
    \node[below left = of yb] (ybsw) {};
    \node[above left = of yb] (ybnw) {};
    \draw[densely dotted] (ybne) --  (ybsw);
    \draw[densely dotted] (ybnw) --  (ybse);
    \end{scope}
  \end{tikzpicture}
\caption{Causal structure for different values of $z$. For simplicity of illustration $\vec{x}$ and $\vec{y}$ are chosen to point into the direction of time. Dotted lines indicate lightcones.}
\label{fig:causal_structure}
\end{figure}
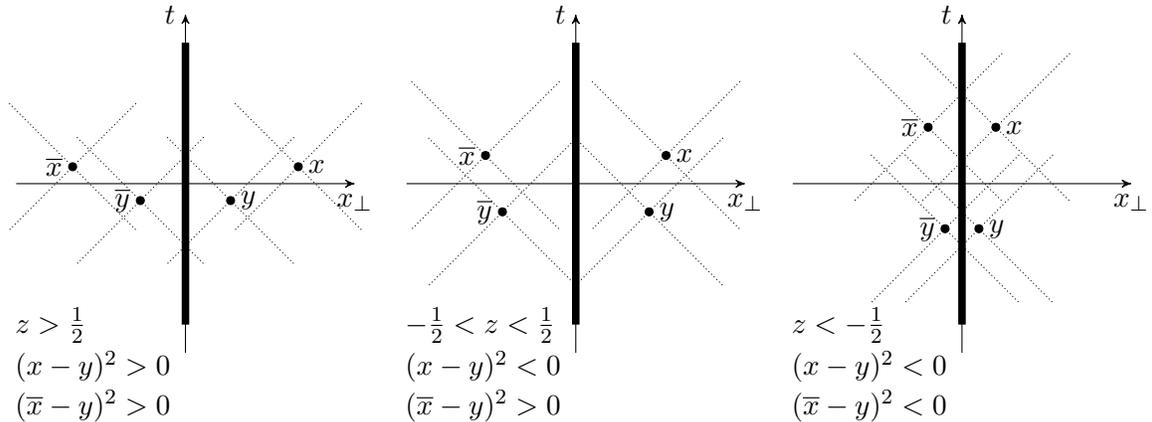

Let us try to understand what the branch cuts mean physically.
In homogeneous CFT in Lorentzian signature, singularities occur in the four point function
whenever an operator crosses the lightcone originating from another operator,
starting from the configuration where all points are spacelike separated and the correlator agrees with the one in Euclidean signature \cite{Hartman:2015lfa}.
We will see that the same thing happens in BCFT.
To this end define the image of the point $x$ under reflection by the boundary $\bar{x} = (\vec{x},-x_\perp)$. The cross-ratio can then be written as
\beq
z= \frac{(x-y)^2}{4x_{\perp} y_{\perp}} + \frac{1}{2} = \frac{(\bar{x}-y)^2}{4x_{\perp} y_{\perp}} - \frac{1}{2}\,,
\eeq
where
\beq
(\bar{x}-y)^2 = (x-y)^2 + 4x_{\perp} y_{\perp} \geq (x-y)^2\,.
\eeq
In Euclidean signature $z$ is restricted to $z \in (\frac{1}{2}, +\infty)$.
When considering Lorentzian signature the distances between $x$, $y$ and its images can become timelike separated. The correlator has singularities at the values of $z$ where an operator crosses the lightcone
originating at the other operator or its image.
For $z < \frac{1}{2}$ the points $x$ and $y$ become timelike separated and then for $z < -\frac{1}{2}$ also $\bar{x}$ and $y$ are timelike separated. This is illustrated in Figure \ref{fig:causal_structure}.
The remaining singularity at $z=\infty$ stems from one of the operators approaching the boundary, regardless of the causal relationships.
The singularity at $z=-\frac{1}{2}$ was also discussed in \cite{Lauria:2017wav}, using Landau diagrams.

\subsection{Image symmetry}

The boundary conformal blocks have an approximate symmetry under the transformation that takes one of the coordinates to its image $x_{\perp} \to - x_{\perp}$ and corresponds to $z \to -z$. 
As we saw above, $-z$ will lie on a branch cut, so we have to analytically continue the blocks along one of the paths shown in Figure \ref{fig:ac} to reach this point.
As indicated in \eqref{eq:branch_cuts} the hypergeometric function has its branch cut at $z \in (-\frac{1}{2}, \frac{1}{2})$, hence we are going around this cut without touching it. The branch cut of the power function in front however makes our result path dependent
\beq
g_{i}(\Delta,z e^{\pm i \pi}) = (z e^{\pm i \pi}-\tfrac{1}{2})^{- \Delta} {}_2 F_1 \left(\Delta,\De + 1 - \frac{d}{2}; 2 \De + 2 - d; \frac{1}{\frac{1}{2} + z}\right) = e^{\mp i \pi \De} g_{i}(\Delta,z)\,.
\label{eq:ac_gi}
\eeq
The result after analytic continuation is unchanged up to a phase or, for integer dimensions, a possible minus sign.
This can be used to remove all boundary blocks with odd/even integer dimensions from the bootstrap equation by
adding/subtracting the bootstrap equation and its analytic continuation.
This image symmetry was already noticed in \cite{McAvity:1995zd} as a symmetry of the correlator at $O(\epsilon)$ in the
epsilon expansion. The reason this correlator has the symmetry is that it is equal to a single boundary block.

\section{Review: BCFT bootstrap up to $O(\epsilon)$}
\label{sec:3}

In this section we will review the boundary bootstrap up to the first order in the epsilon expansion, which was done in \cite{Liendo:2012hy}.
The simplest solution to the bootstrap equation was found in that paper by assuming each side of the bootstrap equation contains only a single conformal block
\beq
\mu_{\hat{\De}} g_i(\hat{\De},z)
=(z-\tfrac{1}{2})^{-\Delta_\phi} 
\left(1 + \l a_{\De} g_b(\Delta,z) \right)\, .
\eeq
The solution to this ansatz is the CFT of a free scalar. The dimensions are fixed to the values
\beq
\De_\phi =  \Delta_\phi^{(0)}  =  \frac{d}{2}-1\,,\qquad
\De =  \Delta_{\phi^2}^{(0)} =  d-2\,,
\label{eq:free_field_Delta}
\eeq
hence $\De$ is the dimension of $\phi^2$, in accordance with the free field bulk OPE
$\phi \times \phi = \mathbf{1} + \phi^2$.
For the remaining coefficients there are two solutions, reproducing the free field results for Neumann and Dirichlet boundary conditions\footnote{We will use the labels $D$ and $N$ for Dirichlet or Neumann boundary conditions respectively,  whenever expressions differ between the two cases.}
\bea
\mu_{\hat{\De}}^N = \mu_{\hat{\phi}}^{(0)N} & =  2   \, ,& & &
\hat{\De}^N = \Delta_{\hat{\phi}}^{(0)N} & =  \frac{d}{2}-1  \, ,& & &
\l a_{\De}^N = \l_{\phi^2}^{(0)N} & =  1  \, , \\
\mu_{\hat{\De}}^D =  \mu_{\hat{\phi}}^{(0)D} & =  \frac{d}{2}-1  \, ,& & &
\hat{\De}^D = \Delta_{\hat{\phi}}^{(0)D} & =  \frac{d}{2}  \, ,& & &
\l a_{\De}^D = \l_{\phi^2}^{(0)D} & =  -1  \, .
\eea{eq:epsilon_ansatz}
The solution shows that as expected, the single operator $\hat{\phi}$ contributing in the boundary channel is $\phi$ for Neumann and $\partial_\perp \phi$ for Dirichlet boundary conditions, i.e.\ has dimension $\De_\phi$ or $\De_\phi+1$. 

After finding this solution, \cite{Liendo:2012hy} went on to find a generalization to interacting CFT by assuming that the CFT data is given by an expansion around the free field values (\ref{eq:free_field_Delta}, \ref{eq:epsilon_ansatz})
and by allowing a finite number of conformal blocks.
In fact it is enough to allow one additional conformal block in the bootstrap equation
\beq
\label{crossingsy}
G_i(z) = G_b(z)\,,
\eeq
where
\bea
G_i(z) &= \mu_{\hat{\phi}} g_i(\Delta_{\hat{\phi}},z)\, ,\\
G_b(z) &= (z-\tfrac{1}{2})^{-\Delta_\phi} 
\left(1 + \l a_{\phi^2} g_b(\Delta_{\phi^2},z) +  \l a_{\phi^4}  g_b(\Delta_{\phi^4},z) \right)\, ,
\eea{eq:G_def}
and
\beq
 \Delta_{\phi^4}^{(0)}  =  2d-4  \, , \quad
 \l_{\phi^4}^{(0)}  = 0 \, .
\eeq
The CFT data is expanded in $\epsilon$, the deviation from 4 dimensions
\beq
d =  4-\e\,,
\eeq
and we adopt the following notation for expanding the dimensions and OPE coefficients
\bea
\Delta_k &= \Delta^{(0)}_k + \Delta^{(1)}_k \e + \Delta^{(2)}_k \e^2 + \ldots \,,\\
\l a_{k} &=\l_{k}^{(0)} + \l_{k}^{(1)} \e  + \l_{k}^{(2)} \e^2  + \ldots\,,\\
\mu_{k} &=\mu_{k}^{(0)} + \mu_{k}^{(1)} \e + \mu_{k}^{(2)} \e^2 + \ldots\,.
\eea{eq:expansion_conventions}
The equation \eqref{crossingsy} can be solved to order $\e$ by expanding around $z = \frac{1}{2}$, with the result
\bea
 \Delta_\phi^{(1)} & =  0 \, , & & &
 \Delta_{\phi^2}^{(1)} & =  2 \a  \, ,& & &
\l_{\phi^2}^{(1)} & =  \a  \, ,& & &
 \l_{\phi^4}^{(1)} & = \frac{\a}{2} \, ,& & &
 \\
 \Delta_{\hat{\phi}}^{(1)} & =  - \a  \, ,& & &
 \mu_{\hat{\phi}}^{(1)N} & =  0 \, ,& & &
 \mu_{\hat{\phi}}^{(1)D} & =  \a \, .
\eea{eq:epsilon_1}
This solution is compatible with the CFT data of the $O(N)$ model
\beq
S = \int d^d x \, \sum\limits_{i=1}^N \left( \frac{1}{2} \left( \partial_\mu \phi^i \right)^2 + \frac{m^2}{2} \phi^i \phi^i +\frac{\lambda}{4!} \left( \phi^i \phi^i \right)^2  \right) \,,
\eeq
at the Wilson-Fisher fixed point, where the coupling is
\beq
\frac{\lambda_{*}}{16 \pi^2} = \frac{3}{N+8} \e + O(\e^2) \,.
\eeq
For this model the coefficient $\alpha$ can be fixed by comparing to the known anomalous dimension of $\phi^2$ \cite{Wilson:1972cf}
\beq
\a = \frac{1}{2}\left( \frac{N+2}{N+8} \right) \, .
\eeq
$\Delta_{\phi^4}^{(1)}$ does not appear in the bootstrap equation at this order because 
$\l a_{\phi^4} $ is of order $\epsilon$.

\section{BCFT bootstrap at $O(\epsilon^2)$}
\label{sec:4}

At order $\epsilon^2$ we expect that an infinite number of new operators enter the bootstrap equation.
In the bulk channel we expect new contributions of the form
\beq
H_b(z) = (z-\tfrac{1}{2})^{-\Delta_\phi} \sum_{n=0}^\infty
\l a_{n} g_{b}(\Delta_n,z)\,.
\label{eq:Hb_def}
\eeq
We make the ansatz that the sum runs over operators with dimensions (and corresponding OPE coefficients)
\beq
\Delta_n = 4 + 2n + O(\e)\,, \qquad \l a_{n} = \l_{n}^{(2)}\e^{2} + O(\e^3)\,.
\eeq
The ansatz can be justified with the standard expansion in Feynman diagrams. Since at the critical point the coupling constant becomes proportional to $\e$, we are dealing with the perturbative expansion. Hence we expect operators of the schematic form (not specifying the correct positions of derivatives) $\Box^{n+1} \phi^2$, $\Box^n \phi^4$, $\Box^{n-1} \phi^6$ or $\Box^{n-2} \phi^8$ which all have dimension $4 + 2n$ and can appear at two loops in the perturbative expansion.\footnote{The reason why $\phi^2$ and $\phi^4$ appear only without derivatives at $O(\e)$ was discussed in \cite{Liendo:2012hy}. We do not know if they do appear with derivatives at order $O(\e^2)$.} Naturally degeneracies can occur and in this case the coefficients $\l a_{n}$ really contain contributions from multiple operators.

In the boundary channel the bare dimensions run over odd numbers for Neumann and even numbers for Dirichlet boundary conditions
\beq
H_i^N(z) = \sum_{\substack{n=1\\n\text{ odd}}}^\infty
\mu^N_{n} g_{i}(n,z)\,, \qquad
H_i^D(z) = \sum_{\substack{n=2\\n\text{ even}}}^\infty
\mu^D_{n} g_{i}(n,z)\,.
\label{eq:Hi_def}
\eeq
The justification is similar to the one in the bulk channel.
For Neumann boundary conditions the exchanged boundary operators are given by the scalar bulk operators which couple to $\phi$, which have odd dimensions. For Dirichlet boundary conditions we expect the normal derivative of these operators, with dimensions increased by one.
The BOE coefficients are of order $\e^2$
\beq
\mu^N_{n} = \mu_{n}^{(2)N}\e^{2} + O(\e^3)\,, \qquad 
\mu^D_{n} = \mu_{n}^{(2)D}\e^{2} + O(\e^3)\,.
\label{eq:OPE_coeffs_boundary}
\eeq
With these assumptions the bootstrap equation at $O(\e^2)$ is
\beq
\label{crossingsy_epsilon2}
 G_i(z) +H_i(z) = G_b(z) +H_b(z) \, ,
\eeq
where $G_i(z)$ and $G_b(z)$ are still given by \eqref{eq:G_def}.
To be more precise about the whereabouts of the CFT data, the functions $G$ include OPE and BOE coefficients up to $O(\e)$ and conformal dimensions up to  $O(\e^2)$, while the $H$ contain the $O(\e^2)$ corrections to the OPE and BOE coefficients. The only exception is the coefficient $\l_{\phi^2}^{(2)}$
which is included in $G_b(z)$ because we will extract the OPE coefficients from $\text{Disc }H_b(z)$, however $\text{Disc }g_{b}(2,z)=0$.
Hence the CFT data included in each function is
\bea
G_i:\qquad & \mu_{\hat \phi}^{(0,1)}, \Delta_{\hat \phi}^{(0,1,2)}\,,  \\
H_i:\qquad & \mu_{n}^{(2)}\,,\\
G_b:\qquad & \l_{\phi^2}^{(0,1,2)}, \l_{\phi^4}^{(1)}, \Delta_{\phi^2}^{(0,1,2)}, \Delta_{\phi^4}^{(0,1)}\,,\\
H_b:\qquad & \l_{n}^{(2)}\,.
\eea{eq:OPE_data_whereabouts}
The functions $G_b(z)$ and $G_i(z)$ to $O(\e^2)$ can be found in Appendix \ref{sec:G}.

\subsection{Bulk OPE coefficients}

As discussed in Section \ref{sec:analytic_structure}, we can take the discontinuity of the bootstrap equation to remove the infinite sum over boundary blocks of integer dimensions $H_i(z)$
\beq
 \text{Disc } G_i(z) = \text{Disc } G_b(z)   +\text{Disc } H_b(z) \, ,
\label{eq:bootstrap_disc}
\eeq
In addition to removing $H_i(z)$, taking the discontinuity also reduces the complexity of the functions appearing. The functions $G_b(z)$ and $G_i(z)$ contain $\text{Li}_2$ and $\log^2$ terms. The discontinuities of these functions however are only simple logarithms. 
Let us start by first
reconstructing the OPE coefficients in $H_b(z)$ and then compute the full function $H_b(z)$
itself. One computes
\bea
\text{Disc } H_b(z) &= \text{Disc } G_i(z) - \text{Disc } G_b(z) \\
&=
\e^2 \pi i \left(
\frac{A + C \log(z-\frac{1}{2}) + D \log(z+\frac{1}{2})}{z + \frac{1}{2}}
+ \frac{B + E \log(z+\frac{1}{2})}{z - \frac{1}{2}}
\right)\,,
\eea{eq:DiscH}
where we defined new coefficients
\bea
A^{N/D} &= 2 \Delta_{\hat{\phi}}^{(2)N/D} - 2 \Delta_{\phi}^{(2)} -\frac{7}{2} \a + 3 \Delta_{\phi^4}^{(1)} \a\,,&&&&\\
B^N &=  2 \Delta_{\hat{\phi}}^{(2)N} - 2 \Delta_{\phi}^{(2)} + \Delta_{\phi^2}^{(2)}-C\,,&&&
B^D &=  -2 \Delta_{\hat{\phi}}^{(2)D} + 2 \Delta_{\phi}^{(2)} - \Delta_{\phi^2}^{(2)}-C\,,\\
C &= (\Delta_{\phi^4}^{(1)} - 2\a - 1) \a\,,&&&
D &= \a-\a\Delta_{\phi^4}^{(1)}\,,&&&&\\
E^N &= -2 \a^2\,, &&& E^D &= 2 \a^2\,.
\eea{eq:ABC}
In order to compute the OPE coefficients, $\text{Disc } H_b(z)$ should be expanded in terms of the discontinuities of conformal blocks
\beq
\text{Disc } H_b(z) =
 \sum_{n=0}^\infty
\l a_{n} \,
\text{Disc} \left(
(z-\tfrac{1}{2})^{-\Delta_\phi}
g_{b}(\Delta_n,z)
\right)\,.
\label{eq:dishH_discBlocks}
\eeq
One might wonder if we are allowed to commute the discontinuity past the infinite sum, especially since it is computed at the boundary of the region of convergence of the bulk OPE, as discussed in Section \ref{sec:analytic_structure}.
Since we will compute $H_b(z)$ below by summing the blocks themselves, this question can be answered in the end by checking that $H_b(z)$ has the correct discontinuity.
That the discontinuity still contains enough information to distinguish the OPE coefficients is ensured by the fact that discontinuities of the conformal blocks to order $\e^0$ are Jacobi polynomials, which are orthogonal
\beq
\text{Disc} \left(
(z-\tfrac{1}{2})^{-\Delta_\phi}
g_{b}(\Delta_n,z)
\right)
= \frac{b_n}{z+\frac{1}{2}}
P_n^{(1,0)} \left(\frac{2}{\frac{1}{2}+z} - 1 \right)\,, \qquad
b_n = - \frac{2 \pi i (-1)^{n} (2n+2)!}{ n! (n+1)!}\,.
\eeq
We can use the orthogonality relation for Jacobi polynomials
\beq
\int_{-1}^1 (1-x) P_n^{(1,0)}(x) P_m^{(1,0)}(x) dx = \frac{4}{2n+2} \delta_{nm}\,,
\eeq
to extract the OPE coefficients
\bea
\l a_{n} ={}& \frac{2n+2}{4 b_n} \int_{-1}^1 (1-x) P_n^{(1,0)}(x)
(z+\tfrac{1}{2}) \text{Disc } H_b(z) 
\Big|_{z \to \frac{2}{1+x} -\frac{1}{2}}\\
={}& \e^2 \pi i  \frac{2n+2}{4 b_n} \int_{-1}^1 P_n^{(1,0)}(x)
\Bigg(
2 B + (1-x) A 
+ \left(2 E + (1-x) D \right) \log \left( \frac{2}{1+x} \right)\\
&+  (1-x) C  \log \left( \frac{2}{1+x} -1\right)
\Bigg)\\
={}&   \frac{2 \pi i \e^2}{b_n}
\begin{cases} \left(  B + \frac{(-1)^n}{n+1+(-1)^n} C + \frac{(-1)^n (n+1)}{n(n+2)} D
+ \frac{(-1)^n}{(n+1)} E
\right)\,, & n > 0\,,\\
\left(\
\frac{1}{2} A + B + \frac{1}{2} C + \frac{3}{4} D + E
\right)
\,, & n = 0\,.
\end{cases}
\eea{eq:inversion_formula}
Knowing the OPE coefficients, we can compute $H_b(z)$ by doing the sum in \eqref{eq:Hb_def}
\bea
H_b(z) ={}& \frac{A\e^2}{2}  \left( \frac{1}{z + \frac{1}{2}} - \frac{\log(z+\frac{1}{2})}{z-\frac{1}{2}} \right)
- \frac{B \e^2}{2} \frac{\log(z+\frac{1}{2})}{z+\frac{1}{2}}
- \frac{C \e^2}{4} \frac{\log(z+\frac{1}{2})^2}{z-\frac{1}{2}}\\
&+ \frac{D \e^2}{2} \left( \frac{1}{z + \frac{1}{2}} + \frac{\text{ Li}_2 \left(\frac{1}{2}-z\right)}{z - \frac{1}{2}}  \right)
+ \frac{E \e^2}{2}  \frac{\text{ Li}_2 \left(\frac{1}{2}-z\right)}{z + \frac{1}{2}}  
\,.
\eea{eq:H_b}
One can now check that the discontinuity of this function is indeed \eqref{eq:DiscH},
which proves that \eqref{eq:dishH_discBlocks} is correct.

\subsection{Using image symmetry}

Given that we expect all the bare dimensions in the boundary channel to be even or odd integers,
the image symmetry of the boundary blocks \eqref{eq:ac_gi} has the following immediate consequences for $H_{i}(z)$
\bea
H^N_{i}(z) + H^N_{i}(z e^{\pm i \pi})  &= O(\epsilon^3)\,,\\
H^D_{i}(z) - H^D_{i}(z e^{\pm i \pi})  &= O(\epsilon^3)\,.
\eea{eq:Hi_vanishing_combinations}
This leads to an (anti-)symmetrized bootstrap equation without $H_i(z)$, depending on the boundary conditions,
\bea
 G^N_i(z) +G^N_i(z e^{\pm i \pi}) &= G^N_b(z)   +G^N_b(z e^{\pm i \pi}) +H^N_b(z)+H^N_b(z e^{\pm i \pi}) \, ,\\
 G^D_i(z) -G^D_i(z e^{\pm i \pi}) &= G^D_b(z)   -G^D_b(z e^{\pm i \pi}) +H^D_b(z)-H^D_b(z e^{\pm i \pi}) \, .
\eea{eq:bootstrap_symmetrized}
Since we already computed $H_b(z)$, everything in these equations is known and we can immediately check if they are satisfied and whether they lead to new constraints on CFT data. 
We find that they are satisfied (for both possible continuation paths) provided that the following relations hold for the CFT data. For Neumann boundary conditions
\beq
\Delta_{\phi^4}^{(1)} = 2 \,, \quad
\Delta_{\hat{\phi}}^{(2)N} = \frac{1}{2}\left(4\Delta_{\phi}^{(2)} - \Delta_{\phi^2}^{(2)} + \a-2 \a^2 \right)\,, \quad
\l_{\phi^2}^{(2)N} = \frac{1}{2}\left(-2\Delta_{\phi}^{(2)} + \Delta_{\phi^2}^{(2)} - \a+2 \a^2 \right) \,,
\label{eq:symmetrized_bootstrap_constraints_N}
\eeq
and for Dirichlet boundary conditions
\beq
\Delta_{\phi^4}^{(1)} = 2 \,, \quad
\Delta_{\hat{\phi}}^{(2)D} = \frac{1}{2}\left(4\Delta_{\phi}^{(2)} - \Delta_{\phi^2}^{(2)} - \a+2 \a^2 \right)\,, \quad
\l_{\phi^2}^{(2)D} =  \frac{1}{2}\left(-2\Delta_{\phi}^{(2)} + \Delta_{\phi^2}^{(2)} + \a-2 \a^2 \right)\,.
\label{eq:symmetrized_bootstrap_constraints_D}
\eeq
The first condition is the correct value for $\Delta_{\phi^4}^{(1)}$ (see e.g.\ \cite{Rychkov:2015naa, Gliozzi:2017hni}), and the other ones can be checked by inserting the known CFT data \cite{Wilson:1972cf}
\beq
\Delta_\phi^{(2)} = \frac{N+2}{4(N+8)^2}\,, \qquad
\Delta_{\phi^2}^{(2)} = \frac{(N+2)(13N+44)}{2(N+8)^3}\,,
\label{eq:2loop_bulk_data}
\eeq
gaining the following anomalous dimension for the boundary operator
\beq
\Delta_{\hat{\phi}}^{(2)N} = - \frac{5(N+2)(N - 4)}{4 (N+8)^3}\,, \qquad
\Delta_{\hat{\phi}}^{(2)D} = - \frac{(N+2)(17 N + 76)}{4 (N+8)^3}\,.
\eeq
These are precisely the values computed for Neumann boundary conditions in \cite{reeve1981} and for 
Dirichlet boundary conditions in \cite{reeve1980,Diehl:1981jgg} (using that $\Delta_{\hat{\phi}} = \frac{1}{2}(d-2 + \eta_{\parallel})$).

Note that these conditions are the only way we can compare to previously known CFT data. Apart from this we only compute previously unknown OPE and BOE coefficients. With the scaling dimensions from this section, we can write down the OPE coefficients given in \eqref{eq:inversion_formula} and (\ref{eq:symmetrized_bootstrap_constraints_D}, \ref{eq:symmetrized_bootstrap_constraints_N}) in terms of $n$ and $N$. For Neumann boundary conditions
\begin{align}
\l_{\phi^2}^{(2)N} &= \frac{3(N + 2)(N - 2)}{2(N + 8)^3} \,, \nonumber\\
\l a_{0}^N &=	\frac{(N+2)(N(N+2)-108)}{8(N+8)^3}\epsilon^2
	\,,\label{eq:bulk_OPE_coeffs_N}\\
\l a_{n>0}^N &= 
	\frac{n! (n+1)!}{2 (2n+2)!}
    \frac{N+2}{(N+8)^2} \left( \frac{6(-1)^n(n+1)}{n+(-1)^n+1} + \frac{N+8}{n(n+1)(n+2)}+ \frac{2(N+5)}{n+1} - 7(-1)^n \right) \epsilon^2\,,\nonumber
\end{align}
and for Dirichlet boundary conditions
\bea
\l_{\phi^2}^{(2)D} &= \frac{3(N + 2)(3N + 14)}{2(N + 8)^3} \,, \\
\l a_{0}^D &= 	-\frac{(N+2) (N (3N + 22) + 44)}{8 (N + 8)^3}\epsilon^2
	\,,\\
\l a_{n>0}^D &= 
	\frac{(-1)^n n! (n+1)!}{2 (2n+2)!}
    \frac{N+2}{(N+8)^2} \left( 1+ \frac{N+8}{n(n+2)} - \frac{N+2}{(n+1)(n+(-1)^n+1)} \right)\epsilon^2
	\,.
\eea{eq:bulk_OPE_coeffs_D}

\subsection{Boundary OE coefficients}

The next step is to consider the full bootstrap equation \eqref{crossingsy_epsilon2} to compute the sum of new boundary blocks and then use an orthogonality relation for the boundary blocks appearing in $H_i(z)$ to compute the corresponding BOE coefficients.
Using the bootstrap equation and \eqref{eq:symmetrized_bootstrap_constraints_N} we have for Neumann boundary conditions
\begin{align}
H^N_i(z) ={}& G^N_b(z) + H^N_b(z) - G^N_i(z)\label{eq:HiN}\\
={}& 
K \e^2 \frac{ \log \left(\frac{z+ \frac{1}{2}}{z- \frac{1}{2}} \right)}{(z+ \frac{1}{2})(z- \frac{1}{2})}
+ L \e^2 \left( \frac{ \text{Li}_2 \left(\frac{1}{\frac{1}{2}-z}\right)}{(z+ \frac{1}{2})(z- \frac{1}{2})} 
+  \frac{\log \left(\frac{z+ \frac{1}{2}}{z- \frac{1}{2}} \right)^2}{2(z- \frac{1}{2})}
  \right)\,,\nonumber
\end{align}
where
\beq
K= \De_{\phi}^{(2)}\,, \qquad
L = \frac{\a(2 \a -1)}{2} \,.
\eeq
For Dirichlet boundary conditions we compute
\begin{align}
H^D_i(z) ={}& 
\frac{J \e^2}{(z+ \frac{1}{2})(z- \frac{1}{2})} +
K \e^2 \frac{2z\log \left(\frac{z+ \frac{1}{2}}{z- \frac{1}{2}} \right)}{(z+ \frac{1}{2})(z- \frac{1}{2})} + L \e^2 \left( \frac{ 2z \text{Li}_2 \left(\frac{1}{\frac{1}{2}-z}\right)}{(z+ \frac{1}{2})(z- \frac{1}{2})} 
+  \frac{\log \left(\frac{z+ \frac{1}{2}}{z- \frac{1}{2}} \right)^2}{2(z- \frac{1}{2})}
  \right)\,,\label{eq:HiD}
\end{align}
with
\beq
J = \frac{1}{2} \left( -4 \De_{\phi}^{(2)} + \De_{\phi^2}^{(2)} - \a +2 \a^2\right)\,.
\eeq
One can check that the functions $H^N_i(z)$ and $H^D_i(z)$ satisfy the conditions
\bea
\text{Disc } H_i(z) &= 0\,,\\
H^N_i(z) + H^N_i(z e^{\pm i \pi}) &= 0\,,\\
H^D_i(z) - H^D_i(z e^{\pm i \pi}) &= 0\,.
\eea{eq:Hi_checks}
This is an important consistency check, as it is required to expand the functions in conformal blocks for odd or even dimensions.

The new blocks in the boundary channel are
\beq
g_{i}(n,z) = \left( z- \frac{1}{2} \right)^{-n}
{}_2 F_1 \left(n,n-1,2n-2,\frac{1}{\frac{1}{2}-z} \right)\,.
\eeq
We can project onto any term from $H_i(z)$ using the orthogonality relation \cite{Heemskerk:2009pn}
\beq
\oint \frac{d x}{2\pi i} x^{n-n'-1} {}_2 F_1(n,n-1,2n-2,-x) {}_2 F_1(1-n',2-n',4-2n',-x) = \delta_{n,n'}\,,
\label{eq:2F1_orthogonality}
\eeq
where the contour circles $0$ counterclockwise. 
Note that the integrated function has an isolated singularity at $x=0$ and a branch cut along the interval $(-\infty,-1)$. When changing to our coordinate $x = \frac{1}{z-\frac{1}{2}}$ the isolated singularity is mapped to $z=\infty$ and the branch cut to $z \in(-\frac{1}{2}, \frac{1}{2})$. Hence this orthogonality relation holds when integrating around the branch cut.
Recall that the discontinuity of $g_{i}(n,z)$ vanishes along this path (Figure \ref{fig:analytic_structure_gi_gb}), so this operation is well defined. 
In practice the easiest way to compute the BOE coefficients is to change variables to $x$ and compute the residue
\beq
\mu_n = \Res\limits_{x=0} \left( x^{-n-1} {}_2 F_1(1-n,2-n,4-2n,-x) H_i \left( \frac{1}{x} + \frac{1}{2}\right)\right)\,.
\label{eq:inversion_boundary}
\eeq
Because of the conditions \eqref{eq:Hi_checks} it is clear that the conformal blocks for odd or even dimensions contribute only to one of the boundary conditions.
We can therefore study both contributions together, in which case the formula simplifies
\beq
H_i^N \left( \tfrac{1}{x} + \tfrac{1}{2}\right) +
H_i^D \left( \tfrac{1}{x} + \tfrac{1}{2}\right) = 
J\e^2 \frac{x^2}{x+1}
+ 2 K \e^2 x \log(1+x)
+   L \e^2 x \left( \log(1+x)^2 +2\text{Li}_2(-x) \right) \,.
\eeq
Inserting this function into \eqref{eq:inversion_boundary}
one finds the BOE coefficients
\beq
\mu^N_{n=1} = 0 \,,
\eeq
and
\beq
\mu^D_{n=2} = (J+2K-2L) \e^2 \,,
\eeq
as well as the remaining BOE coefficients, which are given by the same formula for both boundary conditions, where only the odd or even values of $n$ appear for Neumann or Dirichlet boundary conditions
\beq
\mu^N_{\substack{n \geq 3\\\text{odd}}} = \mu^D_{\substack{n \geq 4\\ \text{even}}} =\e^2 \frac{2^{5-2n} \sqrt{\pi} \Gamma(n-2)}{\Gamma(n-\frac{3}{2})}
\left( K + \frac{(-1)^n }{(n-1)(n-2)} L\right)\,.
\eeq
We can check that these squared BOE coefficients are positive by inserting the known CFT data
\bea
\mu^D_{n=2} &= \frac{(N+2)(19N+92)}{4(N+8)^3} \e^2 \,,\\
\mu^N_{\substack{n \geq 3\\\text{odd}}} &= \frac{2^{5-2n} \sqrt{\pi} \Gamma(n-2)}{\Gamma(n-\frac{3}{2})}
\, \frac{N+2}{(N+8)^2} \,
 \frac{8 + n(n-3)}{4(n-1)(n-2)} \e^2\,,\\
\mu^D_{\substack{n \geq 4\\ \text{even}}} &= \frac{2^{5-2n} \sqrt{\pi} \Gamma(n-2)}{\Gamma(n-\frac{3}{2})}
 \,\frac{N+2}{(N+8)^2} \,
 \frac{(n+1)(n-4)}{4(n-1)(n-2)} \e^2\,.
\eea{eq:boundary_OPE_coeffs_N}

\subsection{Full correlator}
\label{sec:correlator}

The full two point function is given by
\beq
\< \phi (x) \phi(y) \> =\frac{F(z)}{(4 x_\perp y_\perp)^{\Delta_\phi}}
= \frac{G_b(z)+H_b(z)}{(4 x_\perp y_\perp)^{\Delta_\phi}}
=\frac{G_i(z)+H_i(z)}{(4 x_\perp y_\perp)^{\Delta_\phi}}\,,
\eeq
and inserting our results we find\footnote{We thank Vladimir Prochazka and Pedro Liendo for pointing out a typo in earlier versions of this formula.}
\begin{align}
\label{eq:F}
&F^{\pm}(z) = \frac{1}{z-\frac{1}{2}} \pm \frac{1}{z+\frac{1}{2}}
+\left( \frac{\e}{2}  + \frac{(2 \a^2 - \a) \e^2}{12}  \right)\left(\frac{\log \left(z-\frac{1}{2}\right)}{z-\frac{1}{2}}  \pm \frac{\log \left(z+\frac{1}{2}\right)}{z+\frac{1}{2}} \right)\\
&+\left( \a \e + \l_{\f^2}^{(2)\pm} \e^2 \right)\left(\frac{\log \left(z+\frac{1}{2}\right)}{z-\frac{1}{2}}  \pm
\frac{\log \left(z-\frac{1}{2} \right)}{z+\frac{1}{2}} \right)
+\frac{\e^2}{8}\left(\frac{\log^2 \left(z-\frac{1}{2}\right)}{z-\frac{1}{2}}  \pm \frac{\log^2 \left(z+\frac{1}{2}\right)}{z+\frac{1}{2}} \right) \nonumber\\
&+\frac{\a^2 \e^2}{2} \left(\frac{\log^2\left(z+\frac{1}{2}\right)}{z-\frac{1}{2}} \pm
\frac{\log^2\left(z-\frac{1}{2}\right)}{z+\frac{1}{2}}\right)  + \frac{\a \e^2}{2}\left(\frac{1}{z-\frac{1}{2}} \pm \frac{1}{z+\frac{1
   }{2}}\right) \log \left(z+\frac{1}{2}\right) \log
   \left(z-\frac{1}{2}\right)\,,\nonumber
\end{align}
where $F^+$ is the result for Neumann and $F^-$ for Dirichlet boundary conditions.
Note that the final result does not depend on dilogarithms and is totally symmetric or antisymmetric upon exchanging $z-\frac{1}{2}$ with $z+\frac{1}{2}$.

\subsection{Outlook: $O(\epsilon^3)$}

At the next order in $\epsilon$ we can make the ansatz
\beq
\label{crossingsy_epsilon3}
 G_i(z) +H_i(z) +I_i(z) = G_b(z) +H_b(z) +I_b(z) \, ,
\eeq
where the functions $G$ and $H$ are the same as before, but now including OPE coefficients up to $O(\e^2)$ and conformal dimensions up to  $O(\e^3)$.
The new functions $I$ include the corrections to the OPE coefficients at order $\e^3$. They take the role of $H$ in the previous section.
We have in the bulk channel
\beq
I_b(z) = (z-\tfrac{1}{2})^{-\Delta_\phi} \sum_{n=0}^\infty
\e^3 \l^{(3)}_n g_{b}(\Delta_n,z)\,,
\label{eq:Ib_def}
\eeq
and similarly in the boundary channel
\beq
I_i^N(z) = \sum_{\substack{n=1\\n\text{ odd}}}^\infty
\e^3 \mu^{(3)N}_{n} g_{i}(n,z)\,, \qquad
I_i^D(z) = \sum_{\substack{n=2\\n\text{ even}}}^\infty
\e^3 \mu^{(3)D}_{n} g_{i}(n,z)\,.
\label{eq:Ii_def}
\eeq
We can play the same game as before and look at the bootstrap equation for the discontinuity
\beq
 \text{Disc } G_i(z) + \text{Disc } H_i(z) =
 \text{Disc } G_b(z)   +\text{Disc } H_b(z)+\text{Disc } I_b(z)  \, .
\label{eq:bootstrap_disc_epsilon3}
\eeq
We would like to compute $I_b(z)$, which would be equivalent to computing the full two point function to $O(\e^3)$. Let us see what CFT data would be required to achieve this.
\bea
G_i:\qquad &\Delta_{\hat \phi}^{(3)}  \\
H_i:\qquad &{\hat \Delta}_n^{(1)}\\
G_b:\qquad & \Delta_\phi^{(3)}, \Delta_{\phi^2}^{(3)}, \Delta_{\phi^4}^{(2)}\\
H_b:\qquad &\Delta_n^{(1)}
\eea{eq:CFT_data_for_epsilon3}
At least some of the new anomalous dimensions in $G_b$ and $H_b$ are known, see for instance \cite{Rychkov:2015naa}. $\Delta_{\hat \phi}^{(3)}$ is unknown but could result from the bootstrap, similar as it happens at order $\e^2$. 
We are not aware of a result for ${\hat \Delta}_n^{(1)}$ in the literature.

Even if the infinite number of dimensions are known, one has to be careful about possible mixing.
If there is a degeneracy in operator dimensions that is lifted at this order in $\e$, we do not know the individual OPE coefficients since we computed the sum of the OPE coefficients of the degenerate operators at the previous order (as discussed below \eqref{eq:Hb_def}).

\section{Conclusions and outlook}
In this paper we studied the analytic structure of the bootstrap equation arising from two point functions in boundary CFT. Based on this, we proposed a method to extract CFT data from the crossing equations in a perturbative expansion, giving as an input the bulk anomalous dimensions. This approach complements the program presented in \cite{Liendo:2012hy}, and extends \cite{Caron-Huot:2017vep} to the case of boundary CFTs (see also \cite{Lemos:2017vnx} for an extension to higher codimension defects).

It should be possible to apply our method to other theories.
One source of possible examples is the list of solutions to crossing symmetry for scalar operators in  \cite{Liendo:2012hy}.
One could consider the correlator $\< \phi^2 \phi^2 \>$ in the $\e$ expansion which is known to $O(\e)$ \cite{McAvity:1995zd} and try to compute it at the next order. An additional complication is that already the tree level solution requires infinite sums of conformal blocks in both channels \cite{Liendo:2012hy}. It would be interesting to combine the bootstrap conditions coming from this correlator with the ones we obtained in this paper, to further constrain the anomalous dimensions.
 
A very interesting example is the $O(N)$ vector model at large $N$,
which was considered in the context of BCFT in \cite{McAvity:1995zd}.
The tree level correlator is equal to a single conformal block in the boundary channel and expanded in terms of scalar operators of dimensions $2k$, $k \in \mathbb{Z}^+$ in the bulk channel \cite{Liendo:2012hy}. In order to compute the correlator at order $1/N$, our approach would require the anomalous dimensions of these bulk operators as an input.
According to \cite{Lang:1992zw}, the only scalar operators with these dimensions are the powers of the auxiliary field $\l$ that classically replaces $\f^2$ at the large $N$ fixed point and has dimension $d-2\De_\f=2$. The anomalous dimensions for $\l^k$ have been computed in \cite{Lang:1992zw}.

Other examples that would be nice to consider are renormalization group (RG) domain walls, which are systems with an interface between two CFTs which are related by RG flow \cite{Gaiotto:2012np,Gliozzi:2015qsa,Melby-Thompson:2017aip}. 
Under certain circumstances correlators across the interface reproduce the mixing of operators under RG flow. Such systems seem to be suitable to be studied with the method we presented in this paper. 

It would be very interesting also to apply this method to higher codimension defects, see for instance \cite{Lemos:2017vnx}. In this case we expect the same obstacles that we discussed for usual CFTs, meaning that the number of cross ratios is bigger than one, thus the analytic structure will be more complicated. 

Another interesting arena for applying this method is one dimensional CFT, with or without supersymmetry, where similar ideas have already been applied in \cite{Liendo:2018ukf}. In this case there will only be one cross ratio and we expect the method to be feasible. We hope to report on this in the future.

\section*{Acknowledgments}

We thank Fernando Alday, Edoardo Lauria, Madalena Lemos, Marco Meineri, Vladimir Prochazka, and Emilio Trevisani for helpful discussions. In particular, we would like to thank Fernando Alday, Madalena Lemos, Marco Meineri and Vladimir Prochazka for carefully reading and commenting on our first draft.
This research received funding from the Knut and Alice Wallenberg Foundation grant KAW 2016.0129.

\appendix

\section{$G_b$ and $G_i$ at order $\e^2$}
\label{sec:G}

For completeness we list here the expressions for $G_b(z)$ and $G_i(z)$ to order $\e^2$.
The functions generally take the following form, with coefficients $c_j$ that depend on the boundary conditions and whether one considers $G_b(z)$ or $G_i(z)$
\bea
G_{b/i}(z) ={}& \frac{c_1}{z+\frac{1}{2}}+\frac{c_2}{z-\frac{1}{2}}
+\left(\frac{c_3}{z+\frac{1}{2}}+\frac{c_4}{z-\frac{1}{2}}\right) \log \left(z+\frac{1}{2}\right)
+\left(\frac{c_5}{z+\frac{1}{2}}+\frac{c_6}{z-\frac{1}{2}
   }\right) \log
   \left(z-\frac{1}{2}\right)\\
&+\left(\frac{c_7}{z+\frac{1}{2}}+\frac{c_8}{z-\frac{1}{2}
   }\right) \log
   ^2\left(z+\frac{1}{2}\right)
+\left
   (\frac{c_9}{z+\frac{1}{2}}+\frac{c_{10}}{z-\frac{1}{2}}\right) \log
   ^2\left(z-\frac{1}{2}\right)\\
&+\left(\frac{c_{11}}{z+\frac{1}{2}}+\frac{c_{12}}{z-\frac{1
   }{2}}\right) \log \left(z+\frac{1}{2}\right) \log
   \left(z-\frac{1}{2}\right)
+ \left(\frac{c_{13}}{z+\frac{1}{2}}+\frac{c_{14}}{z-\frac{1}{2}}\right)
   \text{Li}_2\left(\frac{1}{2}-z\right)\\
&+ \left(\frac{c_{15}}{z+\frac{1}{2}}+\frac{c_{16}}{z-\frac{1}{2}}\right)
   \text{Li}_2\left(\frac{1}{\frac{1}{2}-z}\right)\,.
\eea{eq:G}
They were computed by expanding \eqref{eq:G_def} in $\epsilon$.
We used the algorithm described in Section 2 of \cite{Huber:2005yg} to expand the hypergeometric functions. The Mathematica package described in that paper cannot be used directly because it assumes the parameters of the hypergeometric functions to be linear in $\e$.
The coefficients are for $G_b(z)$ in Neumann boundary conditions
\begin{equation}
\begin{gathered}
 c_ 1 = 1+  ({\l_{\f^2}^{(1)}}-2 {\l_{\f^4}^{(1)}}) \epsilon+
 \left({\l_{\f^2}^{(2)}}-{\De_{\f^4}^{(1)}} {\l_{
\f^4}^{(1)}}+\frac{1}{2}{\l_{\f^4}^{(1)}}\right) \epsilon ^2
 \,, \qquad
 c_ 2 = 1 \,, \\
 c_ 3 = \frac{\epsilon }{2}  +  ({\l_{\f^2}^{(1)}}-2 
{\l_{\f^4}^{(1)}}) \frac{\epsilon^2}{2}
 \,, \qquad
 c_ 4 = 2 {\l_{\f^4}^{(1)}} \epsilon + \frac{1}{2} (6 {\De_{\f^4}^{(1)}}-7) {\l_{\f^4}^{(1)}} 
\epsilon ^2  \,, \qquad
 c_ 5 = 
 ({\De_{\f^2}^{(1)}}-2 {\De_\f^{(1)}})\frac{\epsilon}{2} +\\
+(-2 
{\De_\f^{(2)}}+{\De_{\f^2}^{(2)}}
-2 {\De_\f^{(1)}} {\l_{\f^2}^{(1)}}
+4 {\De_\f^{(1)}} {\l_{\f^4}^{(1)}}+{\De_{\f^2}^{(1)}} {\l_{\f^2}^{(1)}}
-2 {\De_{\f^4}^{(1)}} {\l_{\f^4}^{(1)}}+2 {\l_{\f^4}^{(1)}}) \frac{\epsilon^2}{2} \,, \\
 c_ 6 = \left(\frac{1}{2}-{\De_\f^{(1)}}\right) \epsilon 
-{\De_\f^{(2)}} \epsilon ^2 
 \,, \qquad
 c_ 7 = \frac{\epsilon ^2}{8} \,, \qquad
 c_ 8 = \frac{{\l_{\f^4}^{(1)}} \epsilon ^2}{2}
 \,, \qquad
 c_ 9 = \frac{1}{8}  ({\De_{\f^2}^{(1)}}-2 
{\De_\f^{(1)}})^2 \epsilon ^2 \,, \\
 c_{10} = \frac{1}{8} (1-2 {\De_\f^{(1)}})^2 \epsilon ^2 
 \,, \qquad
 c_{11} = \frac{1}{4}  ({\De_{\f^2}^{(1)}}-2 
{\De_\f^{(1)}}) \epsilon ^2 \,, \qquad
 c_{12} = {\l_{\f^4}^{(1)}}  (-2 
{\De_\f^{(1)}}+{\De_{\f^4}^{(1)}}-1) \epsilon ^2
 \,, \\
 c_{13} = \frac{{\De_{\f^2}^{(1)}}^2 \epsilon ^2}{4} \,, \qquad
 c_{14} = (2 {\De_{\f^4}^{(1)}}-3) {\l_{\f^4}^{(1)}} \epsilon ^2 
 \,, \qquad
 c_{15} = 
 c_{16} = 0\,,
\end{gathered}
\label{eq:cb_Neumann}
\end{equation}
and in Dirichlet boundary conditions
\begin{equation}
\begin{gathered}
 c_ 1= -1 +  ({\l_{\f^2}^{(1)}}-2 {\l_{\f^4}^{(1)}}) \epsilon
+ \left({\l_{\f^2}^{(2)}}-{\De_{\f^4}^{(1)}} 
{\l_{\f^4}^{(1)}}+\frac{{\l_{\f^4}^{(1)}}}{2}\right) \epsilon^2 \,, \qquad
 c_ 2= 1 \,,\\
 c_ 3= -\frac{\epsilon }{2}+\frac{1}{2}  ({\l_{\f^2}^{(1)}}-2 
{\l_{\f^4}^{(1)}}) \epsilon ^2\,, \qquad
 c_ 4= 2 {\l_{\f^4}^{(1)}} \epsilon + \frac{1}{2} (6 {\De_{\f^4}^{(1)}}-7) {\l_{\f^4}^{(1)}} 
\epsilon ^2 \,, \\
 c_ 5= 
\left(2{\De_\f^{(1)}}-{\De_{\f^2}^{(1)}}\right) \frac{\epsilon}{2}  +
\left({\De_\f^{(2)}}-\frac{{\De_{\f^2}^{(2)}}}{2}
+\left(\frac{1}{2}{\De_{\f^2}^{(1)}}-{\De_\f^{(1)}} \right)
{\l_{\f^2}^{(1)}}
+\left(2 {\De_\f^{(1)}} 
-{\De_{\f^4}^{(1)}} +1\right)
{\l_{\f^4}^{(1)}}\right) \epsilon ^2\,,\\
 c_ 6= \left(\frac{1}{2}-{\De_\f^{(1)}}\right) \epsilon 
-{\De_\f^{(2)}} \epsilon ^2 \,, \qquad
 c_ 7= -\frac{\epsilon ^2}{8} \,, \qquad
 c_ 8= \frac{{\l_{\f^4}^{(1)}} \epsilon ^2}{2} \,, \qquad
 c_ 9=  (2 {\De_\f^{(1)}}-{\De_{\f^2}^{(1)}})^2\frac{\epsilon ^2}{8} \,,\\
 c_{10}= \frac{1}{8} (1-2 {\De_\f^{(1)}})^2 \epsilon ^2 \,, \qquad
 c_{11}= \frac{1}{4}  (2 
{\De_\f^{(1)}}-{\De_{\f^2}^{(1)}})\epsilon ^2 \,, \qquad
 c_{12}= {\l_{\f^4}^{(1)}}  (-2 
{\De_\f^{(1)}}+{\De_{\f^4}^{(1)}}-1) \epsilon ^2\,,\\
 c_{13}= -\frac{1}{4} {\De_{\f^2}^{(1)}}^2 \epsilon ^2 \,, \qquad
 c_{14}= (2 {\De_{\f^4}^{(1)}}-3) {\l_{\f^4}^{(1)}} \epsilon ^2 \,, \qquad
 c_{15}= 
 c_{16} = 0\,.
\end{gathered}
\label{eq:cb_Dirichlet}
\end{equation}
For the function $G_i(z)$ we have for Neumann boundary conditions
\begin{equation}
\begin{gathered}
 c_ 1 =
 c_ 2 = 1+ \frac{{\l_{\hat{\f}}^{(1)}} \epsilon }{2} \,, \qquad
 c_ 3 = \frac{\epsilon}{2} + \frac{{\l_{\hat{\f}}^{(1)}} \epsilon ^2}{4} \,,\qquad
 c_ 4 = 
 c_ 5 = -{\De_{\hat{\f}}^{(1)}} \epsilon
-\left({\De_{\hat{\f}}^{(2)}}+\frac{1}{2}{\De_{\hat{\f}}^{(1)}} {\l_{\hat{
\f}}^{(1)}}\right) \epsilon ^2   \,,\\
 c_ 6 = \frac{\epsilon}{2} + \frac{{\l_{\hat{\f}}^{(1)}} \epsilon ^2}{4} \,, \qquad
 c_ 7 = \frac{\epsilon ^2}{8}  \,, \qquad
 c_ 8 = -\frac{{\De_{\hat{\f}}^{(1)}} \epsilon ^2}{4} \,, \qquad
 c_ 9 = \frac{{\De_{\hat{\f}}^{(1)}}^2 \epsilon ^2}{2}  \,, \\
 c_{10} = \frac{1}{8} \left(-4 {\De_{\hat{\f}}^{(1)}}^2-2 
{\De_{\hat{\f}}^{(1)}}+1\right) \epsilon ^2 \,, \qquad
 c_{11} = -\frac{{\De_{\hat{\f}}^{(1)}} \epsilon ^2}{2}  \,, \qquad
 c_{12} = {\De_{\hat{\f}}^{(1)}}^2 \epsilon ^2 \,, \\
 c_{13} = c_{14} = 0 \,, \qquad
 -c_{15}= 
 c_{16}=-\frac{1}{2} {\De_{\hat{\f}}^{(1)}} (2 {\De_{\hat{\f}}^{(1)}}+1) \epsilon ^2\,,
\end{gathered}
\label{eq:ci_Neumann}
\end{equation}
and for Dirichlet boundary conditions
\begin{equation}
\begin{gathered}
 -c_1 = c_ 2= 1+ ({\De_{\hat{\f}}^{(1)}}+{\l_{\hat{\f}}^{(1)}})\epsilon + \left({\De_{\hat{\f}}^{(2)}}-\frac{1}{2} (2 {\De_{
\hat{\f}}^{(1)}}+1) ({\De_{\hat{\f}}^{(1)}}-{\l_{\hat{\f}}^{(1)}})\right)\epsilon ^2 \,,\\
 c_ 3= -\frac{\epsilon }{2} -\frac{1}{2} 
({\De_{\hat{\f}}^{(1)}}+{\l_{\hat{\f}}^{(1)}})\epsilon ^2 
\,, \qquad
 -c_ 4= 
 c_ 5= {\De_{\hat{\f}}^{(1)}}
 \epsilon +  ({\De_{\hat{\f}}^{(2)}}+{\De_{\hat{\f}}^{(1)}} 
({\De_{\hat{\f}}^{(1)}}+{\l_{\hat{\f}}^{(1)}})) \epsilon ^2\,, \\
 c_ 6= \frac{\epsilon }{2} + \frac{1}{2}  
({\De_{\hat{\f}}^{(1)}}+{\l_{\hat{\f}}^{(1)}})\epsilon ^2 \,, \qquad
 c_ 7= -\frac{\epsilon ^2}{8} \,, \qquad
 c_ 8= -\frac{{\De_{\hat{\f}}^{(1)}} \epsilon ^2}{4} \,, \qquad
 c_ 9= -\frac{1}{2} {\De_{\hat{\f}}^{(1)}}^2 \epsilon ^2 \,,\\
 c_{10}= \frac{1}{8} \left(-4 {\De_{\hat{\f}}^{(1)}}^2-2 
{\De_{\hat{\f}}^{(1)}}+1\right) \epsilon ^2 \,, \qquad
 c_{11}= \frac{{\De_{\hat{\f}}^{(1)}} \epsilon ^2}{2} \,, \qquad
 c_{12}= {\De_{\hat{\f}}^{(1)}}^2 \epsilon ^2 \,,\\
 c_{13}= 
 c_{14}= 0 \,, \qquad
 c_{15}= 
 c_{16}= -\frac{1}{2} {\De_{\hat{\f}}^{(1)}} (2 
{\De_{\hat{\f}}^{(1)}}+1) \epsilon ^2 \,. 
\end{gathered}
\label{eq:ci_Dirichlet}
\end{equation}

\bibliographystyle{JHEP}
\bibliography{boundary_bootstrap}

\providecommand{\href}[2]{#2}\begingroup\raggedright\begin{thebibliography}{10}

\bibitem{Alday:2016njk}
L.~F. Alday, \emph{{Large Spin Perturbation Theory for Conformal Field
  Theories}},
  \href{http://dx.doi.org/10.1103/PhysRevLett.119.111601}{\emph{Phys. Rev.
  Lett.} {\bf 119} (2017) 111601}, [\href{http://arxiv.org/abs/1611.01500}{{\tt
  1611.01500}}].

\bibitem{Caron-Huot:2017vep}
S.~Caron-Huot, \emph{{Analyticity in Spin in Conformal Theories}},
  \href{http://dx.doi.org/10.1007/JHEP09(2017)078}{\emph{JHEP} {\bf 09} (2017)
  078}, [\href{http://arxiv.org/abs/1703.00278}{{\tt 1703.00278}}].

\bibitem{Billo:2016cpy}
M.~Billo, V.~Goncalves, E.~Lauria and M.~Meineri, \emph{{Defects in conformal
  field theory}}, \href{http://dx.doi.org/10.1007/JHEP04(2016)091}{\emph{JHEP}
  {\bf 04} (2016) 091}, [\href{http://arxiv.org/abs/1601.02883}{{\tt
  1601.02883}}].

\bibitem{Gadde:2016fbj}
A.~Gadde, \emph{{Conformal constraints on defects}},
  \href{http://arxiv.org/abs/1602.06354}{{\tt 1602.06354}}.

\bibitem{Liendo:2016ymz}
P.~Liendo and C.~Meneghelli, \emph{{Bootstrap equations for $ \mathcal{N} $ = 4
  SYM with defects}},
  \href{http://dx.doi.org/10.1007/JHEP01(2017)122}{\emph{JHEP} {\bf 01} (2017)
  122}, [\href{http://arxiv.org/abs/1608.05126}{{\tt 1608.05126}}].

\bibitem{deLeeuw:2017dkd}
M.~de~Leeuw, A.~C. Ipsen, C.~Kristjansen, K.~E. Vardinghus and M.~Wilhelm,
  \emph{{Two-point functions in AdS/dCFT and the boundary conformal bootstrap
  equations}}, \href{http://dx.doi.org/10.1007/JHEP08(2017)020}{\emph{JHEP}
  {\bf 08} (2017) 020}, [\href{http://arxiv.org/abs/1705.03898}{{\tt
  1705.03898}}].

\bibitem{Rastelli:2017ecj}
L.~Rastelli and X.~Zhou, \emph{{The Mellin Formalism for Boundary CFT$_d$}},
  \href{http://dx.doi.org/10.1007/JHEP10(2017)146}{\emph{JHEP} {\bf 10} (2017)
  146}, [\href{http://arxiv.org/abs/1705.05362}{{\tt 1705.05362}}].

\bibitem{Soderberg:2017oaa}
A.~Söderberg, \emph{{Anomalous Dimensions in the WF O($N$) Model with a
  Monodromy Line Defect}},
  \href{http://dx.doi.org/10.1007/JHEP03(2018)058}{\emph{JHEP} {\bf 03} (2018)
  058}, [\href{http://arxiv.org/abs/1706.02414}{{\tt 1706.02414}}].

\bibitem{Herzog:2017xha}
C.~P. Herzog and K.-W. Huang, \emph{{Boundary Conformal Field Theory and a
  Boundary Central Charge}},
  \href{http://dx.doi.org/10.1007/JHEP10(2017)189}{\emph{JHEP} {\bf 10} (2017)
  189}, [\href{http://arxiv.org/abs/1707.06224}{{\tt 1707.06224}}].

\bibitem{Karch:2017wgy}
A.~Karch and Y.~Sato, \emph{{Boundary Holographic Witten Diagrams}},
  \href{http://dx.doi.org/10.1007/JHEP09(2017)121}{\emph{JHEP} {\bf 09} (2017)
  121}, [\href{http://arxiv.org/abs/1708.01328}{{\tt 1708.01328}}].

\bibitem{Fukuda:2017cup}
M.~Fukuda, N.~Kobayashi and T.~Nishioka, \emph{{Operator product expansion for
  conformal defects}},
  \href{http://dx.doi.org/10.1007/JHEP01(2018)013}{\emph{JHEP} {\bf 01} (2018)
  013}, [\href{http://arxiv.org/abs/1710.11165}{{\tt 1710.11165}}].

\bibitem{Sato:2017gla}
Y.~Sato, \emph{{More on Boundary Holographic Witten Diagrams}},
  \href{http://dx.doi.org/10.1103/PhysRevD.97.026005}{\emph{Phys. Rev.} {\bf
  D97} (2018) 026005}, [\href{http://arxiv.org/abs/1711.02138}{{\tt
  1711.02138}}].

\bibitem{Lauria:2017wav}
E.~Lauria, M.~Meineri and E.~Trevisani, \emph{{Radial coordinates for defect
  CFTs}},  \href{http://arxiv.org/abs/1712.07668}{{\tt 1712.07668}}.

\bibitem{Lemos:2017vnx}
M.~Lemos, P.~Liendo, M.~Meineri and S.~Sarkar, \emph{{Universality at large
  transverse spin in defect CFT}},  \href{http://arxiv.org/abs/1712.08185}{{\tt
  1712.08185}}.

\bibitem{Goncalves:2018fwx}
V.~Goncalves and G.~Itsios, \emph{{A note on defect Mellin amplitudes}},
  \href{http://arxiv.org/abs/1803.06721}{{\tt 1803.06721}}.

\bibitem{Prochazka:2018bpb}
V.~Prochazka, \emph{{The Conformal Anomaly in bCFT from Momentum Space
  Perspective}},  \href{http://arxiv.org/abs/1804.01974}{{\tt 1804.01974}}.

\bibitem{Bianchi:2018zpb}
L.~Bianchi, M.~Lemos and M.~Meineri, \emph{{Line defects and radiation in
  $\mathcal{N}=2$ theories}},  \href{http://arxiv.org/abs/1805.04111}{{\tt
  1805.04111}}.

\bibitem{Kobayashi:2018okw}
N.~Kobayashi and T.~Nishioka, \emph{{Spinning conformal defects}},
  \href{http://arxiv.org/abs/1805.05967}{{\tt 1805.05967}}.

\bibitem{Karch:2018uft}
A.~Karch and Y.~Sato, \emph{{Conformal Manifolds with Boundaries or Defects}},
  \href{http://dx.doi.org/10.1007/JHEP07(2018)156}{\emph{JHEP} {\bf 07} (2018)
  156}, [\href{http://arxiv.org/abs/1805.10427}{{\tt 1805.10427}}].

\bibitem{Guha:2018snh}
S.~Guha and B.~Nagaraj, \emph{{Correlators of Mixed Symmetry Operators in
  Defect CFTs}},  \href{http://arxiv.org/abs/1805.12341}{{\tt 1805.12341}}.

\bibitem{Isachenkov:2018pef}
M.~Isachenkov, P.~Liendo, Y.~Linke and V.~Schomerus, \emph{{Calogero-Sutherland
  Approach to Defect Blocks}},  \href{http://arxiv.org/abs/1806.09703}{{\tt
  1806.09703}}.

\bibitem{Liendo:2018ukf}
P.~Liendo, C.~Meneghelli and V.~Mitev, \emph{{Bootstrapping the half-BPS line
  defect}},  \href{http://arxiv.org/abs/1806.01862}{{\tt 1806.01862}}.

\bibitem{Lauria:2018klo}
E.~Lauria, M.~Meineri and E.~Trevisani, \emph{{Spinning operators and defects
  in conformal field theory}},  \href{http://arxiv.org/abs/1807.02522}{{\tt
  1807.02522}}.

\bibitem{Liendo:2012hy}
P.~Liendo, L.~Rastelli and B.~C. van Rees, \emph{{The Bootstrap Program for
  Boundary CFT$_d$}},
  \href{http://dx.doi.org/10.1007/JHEP07(2013)113}{\emph{JHEP} {\bf 07} (2013)
  113}, [\href{http://arxiv.org/abs/1210.4258}{{\tt 1210.4258}}].

\bibitem{McAvity:1995zd}
D.~M. McAvity and H.~Osborn, \emph{{Conformal field theories near a boundary in
  general dimensions}},
  \href{http://dx.doi.org/10.1016/0550-3213(95)00476-9}{\emph{Nucl. Phys.} {\bf
  B455} (1995) 522--576}, [\href{http://arxiv.org/abs/cond-mat/9505127}{{\tt
  cond-mat/9505127}}].

\bibitem{Hartman:2015lfa}
T.~Hartman, S.~Jain and S.~Kundu, \emph{{Causality Constraints in Conformal
  Field Theory}}, \href{http://dx.doi.org/10.1007/JHEP05(2016)099}{\emph{JHEP}
  {\bf 05} (2016) 099}, [\href{http://arxiv.org/abs/1509.00014}{{\tt
  1509.00014}}].

\bibitem{Wilson:1972cf}
K.~G. Wilson, \emph{{Quantum field theory models in less than
  four-dimensions}},
  \href{http://dx.doi.org/10.1103/PhysRevD.7.2911}{\emph{Phys. Rev.} {\bf D7}
  (1973) 2911--2926}.

\bibitem{Rychkov:2015naa}
S.~Rychkov and Z.~M. Tan, \emph{{The $\epsilon$-expansion from conformal field
  theory}}, \href{http://dx.doi.org/10.1088/1751-8113/48/29/29FT01}{\emph{J.
  Phys.} {\bf A48} (2015) 29FT01}, [\href{http://arxiv.org/abs/1505.00963}{{\tt
  1505.00963}}].

\bibitem{Gliozzi:2017hni}
F.~Gliozzi, A.~L. Guerrieri, A.~C. Petkou and C.~Wen, \emph{{The analytic
  structure of conformal blocks and the generalized Wilson-Fisher fixed
  points}}, \href{http://dx.doi.org/10.1007/JHEP04(2017)056}{\emph{JHEP} {\bf
  04} (2017) 056}, [\href{http://arxiv.org/abs/1702.03938}{{\tt 1702.03938}}].

\bibitem{reeve1981}
J.~S. Reeve, \emph{Renormalisation group calculation of the critical exponents
  of the special transition in semi-infinite systems},
  \href{http://dx.doi.org/https://doi.org/10.1016/0375-9601(81)90250-4}{\emph{Physics
  Letters A} {\bf 81} (1981) 237 -- 238}.

\bibitem{reeve1980}
J.~S. Reeve and A.~J. Guttmann, \emph{Critical behavior of the $n$-vector model
  with a free surface},
  \href{http://dx.doi.org/10.1103/PhysRevLett.45.1581}{\emph{Phys. Rev. Lett.}
  {\bf 45} (Nov, 1980) 1581--1583}.

\bibitem{Diehl:1981jgg}
H.~W. Diehl and S.~Dietrich, \emph{{Field-theoretical approach to static
  critical phenomena in semi-infinite systems}},
  \href{http://dx.doi.org/10.1007/BF01298293}{\emph{Z. Phys.} {\bf B42} (1981)
  65--86}.

\bibitem{Heemskerk:2009pn}
I.~Heemskerk, J.~Penedones, J.~Polchinski and J.~Sully, \emph{{Holography from
  Conformal Field Theory}},
  \href{http://dx.doi.org/10.1088/1126-6708/2009/10/079}{\emph{JHEP} {\bf 10}
  (2009) 079}, [\href{http://arxiv.org/abs/0907.0151}{{\tt 0907.0151}}].

\bibitem{Lang:1992zw}
K.~Lang and W.~Ruhl, \emph{{The Critical O(N) sigma model at dimensions 2 $<$ d
  $<$ 4: Fusion coefficients and anomalous dimensions}},
  \href{http://dx.doi.org/10.1016/0550-3213(93)90417-N}{\emph{Nucl. Phys.} {\bf
  B400} (1993) 597--623}.

\bibitem{Gaiotto:2012np}
D.~Gaiotto, \emph{{Domain Walls for Two-Dimensional Renormalization Group
  Flows}}, \href{http://dx.doi.org/10.1007/JHEP12(2012)103}{\emph{JHEP} {\bf
  12} (2012) 103}, [\href{http://arxiv.org/abs/1201.0767}{{\tt 1201.0767}}].

\bibitem{Gliozzi:2015qsa}
F.~Gliozzi, P.~Liendo, M.~Meineri and A.~Rago, \emph{{Boundary and Interface
  CFTs from the Conformal Bootstrap}},
  \href{http://dx.doi.org/10.1007/JHEP05(2015)036}{\emph{JHEP} {\bf 05} (2015)
  036}, [\href{http://arxiv.org/abs/1502.07217}{{\tt 1502.07217}}].

\bibitem{Melby-Thompson:2017aip}
C.~Melby-Thompson and C.~Schmidt-Colinet, \emph{{Double Trace Interfaces}},
  \href{http://dx.doi.org/10.1007/JHEP11(2017)110}{\emph{JHEP} {\bf 11} (2017)
  110}, [\href{http://arxiv.org/abs/1707.03418}{{\tt 1707.03418}}].

\bibitem{Huber:2005yg}
T.~Huber and D.~Maitre, \emph{{HypExp: A Mathematica package for expanding
  hypergeometric functions around integer-valued parameters}},
  \href{http://dx.doi.org/10.1016/j.cpc.2006.01.007}{\emph{Comput. Phys.
  Commun.} {\bf 175} (2006) 122--144},
  [\href{http://arxiv.org/abs/hep-ph/0507094}{{\tt hep-ph/0507094}}].

\end{thebibliography}\endgroup
\end{document}